\documentclass[preprint,preprintnumbers,amsmath,amssymb]{revtex4}

\usepackage{graphicx}
\usepackage{dcolumn}
\usepackage{bm}

\newcommand{\be}{\begin{equation}}
\newcommand{\ee}{\end{equation}}
\newcommand{\bea}{\begin{eqnarray}}
\newcommand{\eea}{\end{eqnarray}}

\newcommand{\rme}{{\rm{e}}}

\newcommand{\sig}{{e}}

\begin{document}


\title{N-player quantum games in an EPR setting}
\author{James M. Chappell$^{\text{*}}$, Azhar Iqbal and Derek Abbott}
\affiliation{School of Electrical and Electronic Engineering,
University of Adelaide 5005, Australia \\
$^{\ast }$\texttt{Email}:~\texttt{james.m.chappell@adelaide.edu.au}}


\date{\today}

\begin{abstract}
The $N$-player quantum game is analyzed in the context of an
Einstein-Podolsky-Rosen (EPR) experiment. In this setting, a player's strategies are not unitary transformations as in alternate quantum game-theoretic frameworks, but a classical choice between two directions along which spin or polarization measurements are made. The players' strategies thus remain identical to their strategies in the mixed-strategy version of the classical game. In the EPR setting the quantum game reduces itself to the corresponding classical game when the shared quantum state reaches zero entanglement. We find the relations for the probability distribution for $N$-qubit GHZ and W-type states, subject to general measurement directions, from which
the expressions for the mixed Nash equilibrium and the payoffs are determined.
Players' payoffs are then defined with linear functions so that common two-player games can be easily extended to the $N$-player case and permit analytic expressions for the Nash equilibrium. As a specific example, we solve the Prisoners' Dilemma game for general $ N \ge 2 $. We find a new property for the game that
for an even number of players the payoffs at the Nash equilibrium are equal,
whereas for an odd number of players the cooperating players receive higher payoffs.
\end{abstract}

\maketitle

\section{\label{sec:level1}Introduction}

The field of game theory deals with situations involving strategic interdependence between a set of rational participants.  The study of classical game theory began around 1944 \citep{vonNeumanMorgenstern,Binmore,Rasmusen}, and was extended to the quantum regime, in 1999, by Meyer \citep{MeyerDavid} and Eisert et al \citep{Eisert1999} and has since been developed by many others \cite%
{Blaquiere,Wiesner,Mermin,Peres,Mermin1,Vaidman,BenjaminHayden,EnkPike2002,Johnson,MarinattoWeber,IqbalToor1,DuLi,Du,Piotrowski,IqbalToor3,FlitneyAbbott1,IqbalToor2,Piotrowski1,Shimamura1,FlitneyAbbott2,YongJianHan2002,IqbalWeigert,Mendes,CheonTsutsui,IqbalEpr:2005,Ozdemir2004,CheonAIP,Shimamura,IchikawaTsutsui,OzdemirShimamura,FlitneyGreentree,IqbalCheon,Ichikawa,Ramzan,FlitneyHollenberg2006,Aharon,Bleiler,ahmed2008three,Qiang,NFJP2010,IqbalAbbott,CIL,IqbalCheonAbbott,ChappellB,Chappell3Player,FlitneyAbbottRoyal,NawazToor,GuoZhang,iqbal2009quantum}.
Initially, studies in the arena of quantum games focused on two-player,
two-strategy non-cooperative games but has now been extended to multi-player games by various authors \citep{Popescu1995,BenjaminHayden,iqbalCheonConf,Broom_Cannings_Vickers_1997,QingChen2004,DuLi,du2002Multi,Mermin1,Flitney2009,Boyer2004quant}.  Quantum games have been reported in which players share Greenberger-Horne-Zeilinger
(GHZ) states and W states \citep{YongJianHan2002,Peres,Chappell3Player}, with analysis showing the benefits of players forming coalitions \citep{IqbalToor3,FlitneyGreentree} and also the effects of noise \citep{FlitneyAbbott2,Ramzan}. Such games can be used to describe multipartite situations, such as in the analysis of secure quantum communication \citep{NielsenChuang:2002}.

The usual approach to implementing quantum games involves players sharing a
multi-qubit quantum state with each player having access to an allocated qubit upon which they perform local unitary transformations; then a supervisor submits each qubit to measurement in order to determine the outcome of the game.
An alternative approach in constructing
quantum games uses an Einstein-Podolsky-Rosen (EPR) type setting \citep{IqbalWeigert,IqbalEpr:2005,IqbalCheon,EPR,Bohm,Bell,Bell1,Bell2,Aspect,ClauserShimony,Cereceda,IqbalCheonAbbott}, based on a framework developed by Mermin \cite{Mermin} in 1990. 
In this approach, quantum games are described within an EPR apparatus, with the
players' strategies now being the classical choice between two possible measurement directions implemented when measuring their qubit. This thus becomes equivalent to the standard arrangement for playing a classical mixed-strategy game, in that in each run a
player has a choice between two pure strategies. Thus, as the players'
strategy sets remain classical, the EPR type setting avoids a well known
criticism \citep{EnkPike2002} of conventional quantum games, stemming from the fact that typically, in quantum game frameworks, players are given access to extended strategy sets consisting of local unitary transformations that can be interpreted as fundamentally changing the underlying classical game, and thus not being an authentic extension of it. 

Recently \cite{CIL,ChappellB,Chappell3Player} the formalism of Clifford's geometric algebra
(GA) \citep{Hestenes111,GA,Doran2003,Venzo2007,Dorst:2002} has been
applied in the analysis of quantum games. These works demonstrate
that the formalism of GA facilitates analysis and improves
the geometric visualization of the game.
Multipartite quantum games are usually found significantly harder to
analyze, as we are required to define an $ N \times N $ payoff matrix and calculate measurement outcomes over $ N $-qubit states. In this regard, GA is identified as the most suitable formalism in order to allow
ease of analysis. This becomes particularly convincing in the case where $ N \rightarrow \infty $, where matrix methods become unworkable. As we will later show, an algebraic approach such as GA is both elegant and tractable as $ N \rightarrow \infty $.

Using an EPR type setting we firstly determine the probability distribution of measurement outcomes, giving the player payoffs, and then determine constraints that ensure a faithful embedding of the mixed-strategy version of the original classical game
within the corresponding quantum game.  We then apply our results to an $ N $ player prisoner dilemma (PD) game.

\section{EPR setting for playing multi-player quantum games}

\begin{figure}[htb]
\begin{center}
\includegraphics[width=3.5in]{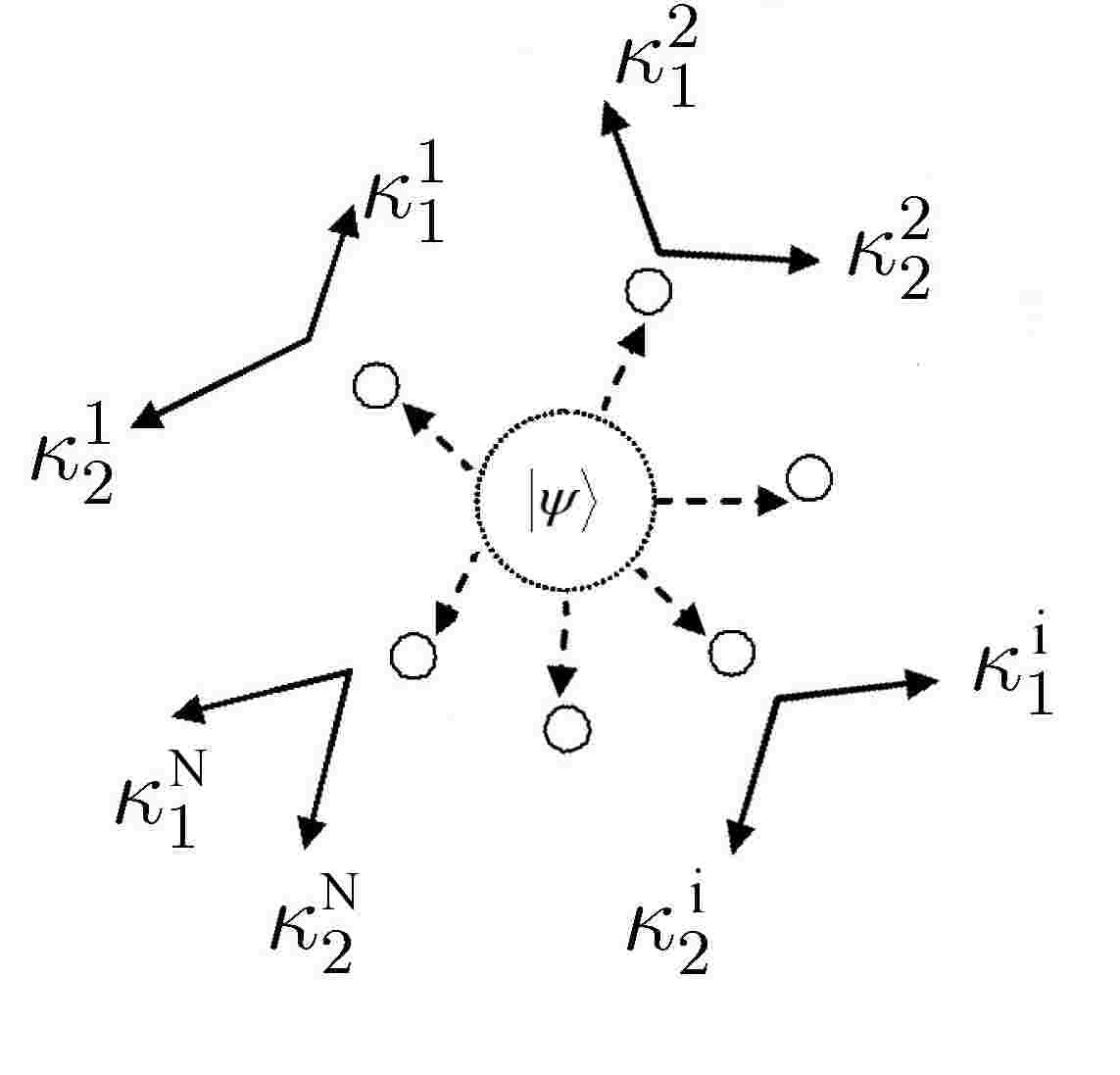}
\end{center}
\caption{The EPR setup for an $N$-player quantum game. In this setup, each player $ i $ has a choice of two measurement directions  $ \kappa _{1}^{i} $ and $ \kappa _{2}^{i}$ for their qubit, allocated from a shared $N$-qubit quantum state.}
\label{eprFigure}
\end{figure}

The EPR setting \citep{IqbalWeigert,IqbalCheon,IqbalCheonAbbott} for a
multi-player quantum game assumes that players $P^i$ are
spatially-separated participants of a non-cooperative game, who are located
at the $ N $ arms of an EPR system \citep{Peres}, as shown in Fig.~\ref{eprFigure}. 
In one run of the experiment, each player chooses one out of
two possible measurement directions. These two directions in space, along which spin or
polarization measurements can be made, are the players' strategies.
As shown in Fig.~\ref{eprFigure}, we represent the $ i^{\rm{th}} $ players' two measurement directions as $%
\kappa _{1}^{i},\kappa _{2}^{i}$, with a measurement returning $+1$ or $-1$.

Over a large number of runs consisting of a sequence of $N$-particle
quantum systems emitted from a source, upon which measurements are performed on each qubit, subject to the players choices of measurement direction, a record is maintained of the experimental outcomes from which players' payoffs can be determined. These payoffs depend on the $N$-tuples of the various 
players' strategic choices made over a large number of runs and on the dichotomic outcomes (measuring spin-up or spin-down) from the measurements performed along those directions.

\subsection{\label{GAintro} Clifford's geometric algebra (GA) }

Typically in a quantum game analysis the tensor product formalism along with Pauli matrices are employed, however matrices become cumbersome for higher dimensional spaces, and so GA is seen as an essential substitute in this case, where the tensor product is replaced with the geometric product and the Pauli matrices are replaced with algebraic elements. The use of GA has also previously been developed in the context of quantum information processing \citep{GA4}.  

To setup the required algebraic framework, we firstly denote $\{\sig _{i}\}$ as a basis for $\Re ^{3}$. Following \cite{ChappellB,Chappell3Player}, we can then form the bivectors $ e_i e_j $, which are non-commuting for $ i \ne j $, with $ e_i e_j = - e_j e_i $ but if $ i = j $ we have  $ e_i^2 = e_i e_i = 1 $. 
We also have the trivector 
\begin{equation}
\iota =\sig _{1}\sig _{2}\sig _{3},
\end{equation}%
finding $\iota ^{2}=\sig _{1}\sig _{2}\sig
_{3}\sig _{1}\sig _{2}\sig _{3}=-1$ and furthermore, that $\iota $
commutes with each vector $\sig _{i}$, thus acting in a similar fashion to the unit imaginary $ \sqrt{-1} $.
We have $\sig _{1}\sig _{2}=\sig _{1}\sig _{2}\sig _{3}\sig
_{3}=\iota \sig _{3}$ and so $\sig _{i}\sig _{j}=\iota \sig _{k}$
for cyclic $i,j,k$.  We can therefore summarize the algebra of the
basis elements $\{\sig _{i}\}$ by the relation 
\begin{equation}
\sig _{i}\sig _{j}=\delta _{ij}+\iota \epsilon _{ijk}\sig _{k},
\label{eq:pauliGA}
\end{equation}%
which is isomorphic to  the algebra of the Pauli matrices \citep{Doran2003}, but now defined as part of $\Re ^{3}$.  

In order to express quantum states in GA we use the one-to-one mapping \citep{Doran2003,Dorst:2002} defined as follows 
\begin{equation}
|\psi \rangle =\alpha |0\rangle +\beta |1\rangle =%
\begin{bmatrix}
a_{0}+\mathrm{i}a_{3} \\ 
-a_{2}+\mathrm{i}a_{1}%
\end{bmatrix}%
\leftrightarrow \psi =a_{0}+a_{1}\iota \sig _{1}+a_{2}\iota \sig
_{2}+a_{3}\iota \sig _{3},  \label{eq:spinorMapping}
\end{equation}%
where $a_{i}$ are real scalars and $ \mathrm{i} = \sqrt{-1} $.

\subsection{\label{sharedStates} Symmetrical $ N $ qubit states}

For $N$-player quantum games an entangled state of $ N $ qubits is prepared, which for fair games should be symmetric with regard to the interchange of the $N$ players, and it is assumed that all information about the state once prepared is known by the players.  
Two types of entangled starting states can readily be identified which are symmetrical with respect to the $ N $ players.
The GHZ-type state 
\be
| {\rm{GHZ}} \rangle_N = \cos \frac{\gamma}{2} |00 \dots 0 \rangle  + \sin \frac{\gamma}{2} |11 \dots 1 \rangle, 
\ee
where we include an entanglement angle $ \gamma \in [-\frac{\pi}{2},\frac{\pi}{2}] $ and the $ W $-type state 
\be
| {\rm{W}} \rangle_N = \frac{1}{\sqrt{N}}\left (|1000 \dots 00 \rangle + |0100 \dots 00 \rangle + |0010 \dots 00 \rangle + \dots + |0000 \dots 01 \rangle \right ).
\ee
To represent these in geometric algebra, we start with the mapping for a single qubit from Eq.~(\ref{eq:spinorMapping}), finding
\be
|0\rangle  \longleftrightarrow  1 , \,\, |1\rangle  \longleftrightarrow  -\iota \sig_2 , 
\ee
so that for the GHZ-type state in GA we have
\be
\psi_{{\rm{GHZ}}_N} = \cos \frac{\gamma}{2} + (-)^N \sin \frac{\gamma}{2} \iota \sig_2^1 \iota \sig_2^2 \dots \iota \sig_2^N,
\ee
where the superscript on each bivector indicates which particle space it refers to. Also for the W-type state we have in GA
\be \label{eq:psiWGA}
\psi_{{\rm{W}}_N} = -\frac{1}{\sqrt{N}} \left (\iota \sig_2^1 + \iota \sig_2^2 + \dots + \iota \sig_2^N \right ).
\ee

\subsection{\label{generalUnitary} Unitary operations and observables in GA}

General unitary operations on a single qubit in GA can be represented as
\be \label{rotorDefn}
R(\theta_1,\theta_2,\theta_3) = \rme^{-\theta_3 \iota \sig_3/2} \rme^{-\theta_1 \iota \sig_2/2} \rme^{-\theta_2 \iota \sig_3/2},
\ee
which is the Euler angle form of a rotation that can completely explore the space of a single qubit, and is equivalent to a general local unitary transformation.
We define $ U^i = R(\theta_1^i,\theta_2^i,\theta_3^i) $ for a general unitary transformation acting locally on each qubit $ i $, which the supervisor applies to the individual qubits that gives the starting state 
\be \label{eq:EisertFinal}
\left (U^1 \otimes U^2 \otimes \dots \otimes U^N \right ) |\psi \rangle,
\ee
upon which the players now decide upon their measurement directions.

The overlap probability between two states $ \psi $ and $ \phi $, in the $N$-particle case \citep{Doran2003}, is
\be \label{NParticleProbabilities}
P(\psi,\phi) = 2^{N-2} \langle \psi E \psi^{\dagger} \phi E \phi^{\dagger} \rangle_0 - 2^{N-2} \langle \psi J \psi^{\dagger} \phi J \phi^{\dagger} \rangle_0 ,
\ee
where the angle bracket $ \langle \,\, \rangle_0 $ indicates that we retain only the scalar part of the product, and where 
\bea \label{EObservable}
E & = & \prod_{b=2}^N \frac{1}{2}(1 - \iota \sig_3^1 \iota \sig_3^b) =  \frac{1}{2^{N-1}}  \left (1 +  \sum_{r=1}^{\lfloor \frac{N}{2}\rfloor}  (-)^r C_{2r}^N (\iota \sig_3^i) \right ), \\ \nonumber
\eea
where $ \lfloor x \rfloor $ returns the nearest integer less than or equal to a given number $ x $, and where we define $ C_r^N(\iota \sig_3^i) $ to represent all possible combinations of $ N $ items taken $ r $ at a time, acting on the object inside the bracket. For example $ C_2^3(\iota \sig_3^i) = \iota \sig_3^1 \iota \sig_3^2 + \iota \sig_3^1 \iota \sig_3^3 + \iota \sig_3^2 \iota \sig_3^3 $. The number of terms produced being given by the standard combinatorial formula $ C_r^N = \frac{N!}{r! (N-r)!} $.

We also have 
\be
J = E \iota \sig_3^1 = \frac{1}{2^{N-1}}  \sum_{r=1}^{\lfloor  \frac{N+1}{2} \rfloor }  (-)^{r+1} C_{2r-1}^N (\iota \sig_3^i), 
\ee
where for simplicity, we initially assume that $ N $ is odd, which simplifies our derivation, and our results can easily be generalized later for all $ N $.  

The supervisor now submits each qubit for measurement, through $ N $ Stern-Gerlach type detectors, with each detector being set at one of the two angles chosen by each player.  As mentioned, each player's choice, is a classical choice between two possible measurement directions, and hence each player's strategy set remains the same as in the classical game, with the quantum outcomes arising solely from the shared quantum state.

In order to calculate the measurement outcomes, we define a separable state $ \phi = A_1 A_2 \dots A_N $, to represent the players directions of measurement, where $ A_i $ is a rotor defined in Eq.~(\ref{rotorDefn}), with probabilistic outcomes calculated according to Eq.~(\ref{NParticleProbabilities}). The use of Eq.~(\ref{NParticleProbabilities}) gives the projection of the state $ \psi $ onto $ \phi $, and thus returns identical quantum mechanical probabilities conventionally calculated using the projection postulate of quantum mechanics. The set of $ | 0 \rangle $ and $ | 1 \rangle $ outcomes obtained from the measurement of each of the $ N $ qubits gives a reward to each player $ p $ according to a payoff matrix $ G^p $.  The expected payoff for each player then calculated from
\be \label{eq:AlicePayoff}
\Pi_p = \sum_{i^1, \dots ,i^N=0}^1 G_{i^1 \dots i^N}^p P_{i^1 \dots i^N} = f(P_{i^1 \dots i^N}) ,
\ee
where $ P_{i^1 \dots i^N} $ is the probability of recording the state $ |i^1 \rangle |i^2 \rangle \dots |i^N\rangle$ upon measurement, where $ i^1,\dots ,i^N \in \{0,1\} $, and $ G_{i^1 \dots i^N}^p $ is the payoff for this measured state. For large $ N $ it is preferable to calculate the payoff as some function $ f $ of the measured states, to avoid the need for large $ N \times N $ payoff matrices, as developed in Section E.2. 

\subsection{GHZ-type state}

Firstly, we calculate the probability distribution of measurement outcomes from Eq.~(\ref{NParticleProbabilities}), from which we then calculate player payoffs from Eq.~(\ref{eq:AlicePayoff}).
For the GHZ-type state we have the first observable given by Eq.~(\ref{EObservable}) producing
\bea \label{EObservableCalc}
\psi E \psi^{\dagger} 
& = & \frac{1}{2^{N-1}} \left ( \prod_{i=1}^N U^i \right ) \left (1 +  \sum_{r=1}^{\lfloor \frac{N}{2} \rfloor }  (-)^r C_{2r}^N (\iota \sig_3^i) \right ) \left ( \prod_{i=1}^N U^{i^{\dagger}} \right ) \\ \nonumber
& = & \frac{1}{2^{N-1}} \left (1 +  \sum_{r=1}^{\lfloor \frac{N}{2} \rfloor }  (-)^r C_{2r}^N (V_3^i) \right ), \nonumber
\eea
where we define $ V_k^j = \iota U^j \sig_k U^{j^{\dagger}} $, and
\be \label{JObservableCalc}
\psi J \psi^{\dagger} = \frac{1}{2^{N-1}} \cos \gamma \sum_{r=1}^{\lfloor \frac{N+1}{2} \rfloor }  (-)^{r+1} C_{2r-1}^N (V_3^i)  - \sin \gamma \Big (\sum_{r=0}^{\lfloor N/2 \rfloor} (-)^{r+\frac{N-1}{2} } C_{2r}^N ( V_2^i V_2^j) V_1^k \dots V_1^N \Big ).  
\ee
For the measurement settings with a separable wave function $ \phi = \prod_i A^i $, we deduce the observables by setting $ \gamma = 0 $ in Eq.~(\ref{EObservableCalc}) and Eq.~(\ref{JObservableCalc}) to be
\bea
\phi J \phi^{\dagger} & = & \frac{1}{2^{N-1}}  \sum_{r=1}^{\lfloor \frac{N+1}{2} \rfloor }  (-)^{r+1} C_{2r-1}^N (M_3^i) \\ \nonumber
\phi E \phi^{\dagger} & = & \frac{1}{2^{N-1}} \left (1 +  \sum_{r=1}^{\lfloor \frac{N}{2} \rfloor }  (-)^r C_{2r}^N (M_3^i) \right ), \nonumber
\eea
where $ M_k^j = \iota A^j \sig_k A^{j^{\dagger}} $.
For $ A^j = e^{-\iota \kappa \sig_2^j/2} $ that allows a rotation of the detectors by an angle $ \kappa $, we find
\bea \label{MeasurementObs}
\phi J \phi^{\dagger} & = & \frac{1}{2^{N-1}}  \sum_{r=1}^{\lfloor \frac{N+1}{2} \rfloor }  (-)^{r+1} C_{2r-1}^N \left ( \iota \sig_3^i e^{\iota \kappa \sig_2^i} \right ) \\ \nonumber
\phi E \phi^{\dagger} & = & \frac{1}{2^{N-1}} \left (1 +  \sum_{r=1}^{\lfloor \frac{N}{2} \rfloor }  (-)^r C_{2r}^N \left ( \iota \sig_3^i e^{\iota \kappa \sig_2^i} \right )  \right ).  \nonumber
\eea
It should be noted in Eq.~(\ref{MeasurementObs}) that we have defined the measurement angles with a simplified rotor, $ e^{-\iota \kappa \sig_2^i/2} $, and we assume no loss of generality, which is in accordance with the known result \citep{Peres} that Bell's inequalities
can still be maximally violated when the allowed directions of measurement are located in a single plane, as opposed to being
defined in three dimensions.

So, referring to Eq.~(\ref{NParticleProbabilities}), we find, through combining Eq.~(\ref{EObservableCalc}) and Eq.~(\ref{MeasurementObs})
\bea \label{eq:Method1MeasureE}
2^{N-2} \langle \psi E \psi^{\dagger} \phi E \phi^{\dagger} \rangle_0 & = & \frac{1}{2^N} \Big \langle \Big (1 +  \sum_{r=1}^{\lfloor \frac{N}{2} \rfloor }  (-)^r C_{2r}^N (V_3^i) \Big )  \Big (1 +  \sum_{r=1}^{\lfloor \frac{N}{2} \rfloor }  (-)^r C_{2r}^N (  \iota \sig_3^i e^{\iota \kappa \sig_2^i}) \Big ) \Big  \rangle_0 \\ \nonumber
& = & \frac{1}{2^N} \left (1+\sum_{r=1}^{\lfloor \frac{N}{2} \rfloor } C_{2r}^N (K^i ) \right ),  \nonumber
\eea
where $ K^i = V_3^i \iota e_3^i \rme^{\iota \kappa e_2^i } = \cos \kappa^i \cos \alpha_1^i + \sin \kappa^i \sin \alpha_1^i \cos \alpha_3^i $, using the standard results listed in Appendix~A.
The cross terms in the expansion of the brackets in Eq.~(\ref{eq:Method1MeasureE}), do not contribute because we only  retain the scalar components in this expression.
We also have for the second part of Eq.~(\ref{NParticleProbabilities}), through combining Eq.~(\ref{JObservableCalc}) and Eq.~(\ref{MeasurementObs})
\be \label{eq:Method1MeasureJ}
 - 2^{N-2} \langle \psi J \psi^{\dagger} \phi J \phi^{\dagger} \rangle_0 = \frac{1}{2^N} \Big (\cos \gamma \sum_{r=1}^{\lfloor \frac{N+1}{2} \rfloor }  C_{2r-1}^N (K^{i} )  +  \sin \gamma \Omega \Big ),  
\ee
where we define
\bea \label{Xidefn}
\Omega & = &  \sum_{r=0}^{\lfloor N/2 \rfloor} (-)^{r} C_{2r}^N ( X_2^i X_2^j) X_1^k \dots X_1^N \\ \nonumber
X_1^i & = & V_1^i \iota e_3^i \rme^{\iota \kappa e_2^i } = \left (-\sin \kappa (\cos \alpha
_{1}\cos \alpha _{2}\cos \alpha _{3}-\sin \alpha _{2}\sin \alpha _{3})+\sin
\alpha _{1}\cos \alpha _{2}\cos \kappa \right )^i  \\ \nonumber
X_2^i & = & V_2^i \iota e_3^i \rme^{\iota \kappa e_2^i } =  \left ( \sin \kappa (\cos \alpha
_{2}\sin \alpha _{3}+\sin \alpha _{2}\cos \alpha _{3}\cos \alpha _{1})-\sin
\alpha _{1}\sin \alpha _{2}\cos \kappa \right )^i ,
\eea
also referring to Appendix A.

\subsubsection{Probability amplitudes for $N$ qubit state, general measurement directions}

So combining our last two results from Eq.~(\ref{eq:Method1MeasureE}) and Eq.~(\ref{eq:Method1MeasureJ}) using Eq.~(\ref{NParticleProbabilities}), we find the probability to find any outcome after measurement, which can be shown to be valid for all $ N $ not just $ N $ odd as initially assumed, is
\be \label{eq:finalDensityGeneral}
P_{k^1 \dots k^N}  =  \frac{1}{2^N} \Big (1+\sum_{r=1}^{\lfloor \frac{N}{2} \rfloor } C_{2r}^N ( \epsilon^{i} K^{i} ) + \cos \gamma \sum_{r=1}^{ \lfloor \frac{N+1}{2} \rfloor } C_{2r-1}^N (\epsilon^{i} K^{i} )   + \epsilon^{1 \dots N}  \Omega \sin \gamma  \Big ) , 
\ee
where we have included $ \epsilon^i = (-)^{k^i} \in \{+1,-1\} $, to select the probability to measure spin-up or spin-down on a given qubit. 

If we take $ \gamma = 0 $, describing the classical limit, we have from Eq.~(\ref{eq:finalDensityGeneral})
\bea
P_{k^1 \dots k^N} & = & \frac{1}{2^N} \left (1+\sum_{r=1}^{\left\lfloor N/2 \right \rfloor} C_{2r}^N (\epsilon^{i} K^{i} ) +  \sum_{r=1}^{\left\lfloor (N+1)/2 \right \rfloor} C_{2r-1}^N (\epsilon^{i} K^i ) \right ) \\ \nonumber
 & = & \frac{1}{2^N} \left (1+\sum_{r=1}^{ N} C_{r}^N ( \epsilon^{i} K^i) \right )  \\ \nonumber
 & = & \frac{1}{2^N} (1 + \epsilon^{1} K^1)(1 + \epsilon^{2} K^2) \dots (1 +\epsilon^{N} K^N), \nonumber
\eea
which shows that for zero entanglement we can form a product state as expected.
Alternatively  with general entanglement, but only for operations on the first two qubits, we have
\be \label{eq:twoQubits}
P_{k^i k^j}  = \frac{1}{8} \left (1 + \epsilon^{k} \cos \gamma \right ) \left (1+ \sum_{r=2}^{ N} C_{r}^N ( \epsilon^{i}) \right ) \left (1 + \epsilon^{ik} K^i)(1 + \epsilon^{jk} K^j \right ), 
\ee
which shows that for the GHZ-type entanglement that each pair of qubits is mutually un-entangled, a well-known result for GHZ-type states.

\subsubsection{Player payoffs}

In general, to represent the permutation of signs introduced by the measurement operator we can define for the first player, say Alice,
\be \label{aDefn}
a^{i^1 \dots i^N} = \frac{1}{2^N} \sum_{j^1 \dots j^N=0}^1 \epsilon^{i^1 \dots i^N} G_{j^1 \dots j^N}^1,  
\ee
so for example, $ a^{0 \dots 0}  =  \frac{1}{2^N} \sum_{j^1 \dots j^N=0}^1 G_{j^1 \dots j^N}^1 $, and we adopt the notation $ a^i a^j = a^{ij} $ etc., i.e.~we write $ a^{0 \dots 1 \dots 0} $ with a 1 in the $ i $th position as $ a^i $. 

Using the payoff function we find for Alice
\be
\Pi_A(\kappa_j^i) = a^{0 \dots 0}+ \sum_{r=1}^{\left\lfloor N/2 \right \rfloor} C_{2r}^N (a^{i} K^i)  + \cos \gamma \sum_{r=1}^{\left\lfloor (N+1)/2 \right \rfloor} C_{2r-1}^N( a^{i} K^i) + a^{k^1 \dots k^N} \Omega  \sin \gamma  
\ee
and similarly for the second player, say Bob, where we would use Bob's payoff matrix in place of Alice's.

\subsubsection{Mixed-strategy payoff relations}

For a mixed strategy game, players choose their first
measurement direction $\kappa _{1}^{i},$  with probabilities $x^i $, where $x^i \in
\lbrack 0,1]$ and hence choose the direction $\kappa _{2}^{i} $ with probabilities $(1-x^i)$,
respectively. Then Alice's payoff is now given as 
\begin{eqnarray}
&&\Pi _{A}(x^1,x^2, \dots,x^N)  \notag \\
&=&x^1 \dots x^N \sum_{i,j,k=0}^{1}P_{i^1 \dots i^N}(\kappa _{1}^{1},\kappa _{1}^{2},\dots ,\kappa
_{1}^{3})G_{i^1 \dots i^N} \\ \nonumber
&+ & \dots + x^1 (1-x^2) \dots x^N \sum_{i,j,k=0}^{1}P_{i^1 \dots i^N}(\kappa _{1}^{1},\kappa
_{2}^{2}, \dots ,\kappa _{1}^{3})G_{i^1 \dots i^N}  \notag \\
&+& \dots + (1-x^1)(1-x^2) x^3 \dots x^N \sum_{i,j,k=0}^{1}P_{i^1 \dots i^N}(\kappa _{2}^{1},\kappa
_{2}^{2},\kappa
_{1}^{3}, \dots , \kappa _{1}^{N})G_{i^1 \dots i^N}  \notag \\
& + & \dots +  (1-x^1)(1-x^2)(1-x^3) \dots (1-x^N) \sum_{i,j,k=0}^{1}P_{i^1 \dots i^N}(\kappa
_{2}^{1},\kappa _{2}^{2},\kappa _{2}^{3}, \dots , \kappa_2^N)G_{i^1 \dots i^N}.  \notag \\
&&  \label{eq:AlicePayoffFourCoin}
\end{eqnarray}

\subsection{Embedding the classical game}

If we consider a strategy $N$-tuple $(x^1,x^2,x^3,\dots,x^N)=(0,1,0, \dots 0)$ for example, at zero
entanglement, then the payoff for Alice is obtained from Eq.~(\ref{eq:AlicePayoffFourCoin}) to be 
\begin{eqnarray}
\Pi _{A}(x^1,\dots ,x^N) &=&\frac{1}{2^N}%
[G_{000 \dots 0}(1+K_{2}^1)(1+K_{1}^2)(1+K_{2}^3) \dots (1+K_{2}^N) \\ \nonumber 
&  & + G_{100 \dots 0}(1-K_{2}^1)(1+K_{1}^2)(1+K_{2}^3) \dots (1+K_{2}^N) 
\notag \\
&&+G_{010 \dots 0}(1+K_{2}^1)(1-K_{1}^2)(1+K_{2}^3) \dots (1+K_{2}^N) \\ \nonumber
& & +G_{110 \dots 0}(1-K_{2}^1)(1-K_{1}^2)(1+K_2^3) \dots (1+K_{2}^N) 
\notag \\
& & + \dots +G_{111 \dots 1}(1-K_{2}^1)(1-K_{1}^2)(1-K_{2}^3) \dots (1-K_{2}^N)].
\end{eqnarray}%
Hence, in order to achieve the classical payoff of $G_{101 \dots 1}$, we can see that we require $K_{2}^1=-1$, $K_{1}^2=+1$ and $K_{2}^3 \dots K_{2}^N=-1$.

This shows that we can select any required classical payoff by the
appropriate selection of $K_{j}^i =\pm 1$. We therefore have the conditions for obtaining
the classical mixed-strategy payoff relations as 
\be
K_{j}^i = \cos \alpha _{1}^i \cos \kappa _{j}^{i}+\sin \alpha_{1}^i \cos \alpha
_{3}^i \sin \kappa_{j}^{i}=\pm 1. 
\ee

We find two classes of solution: If $\alpha
_{3}^i \neq 0$, then for the equations satisfying $K_{2}^i=-1$ we
have for Alice in the first equation $\alpha_{1}^i =0$, $\kappa _{2}^{i}=\pi $
or $\alpha _{1}^i=\pi $, $\kappa _{2}^{i}=0$ and for the equations satisfying $%
K_{1}^i=+1$ we have $\alpha_{1}^i =\kappa_{1}^{i}=0$ or $\alpha
_{1}^i=\kappa_{1}^{i}=\pi $, which can be combined to give either $\alpha
_{1}^i=0$, $\kappa _{1}^{i}=0$ and $\kappa _{2}^{i}=\pi $ or $\alpha _{1}^i=\pi $%
, $\kappa _{1}^{i}=\pi $ and $\kappa _{2}^{i}=0$. For the second class with $%
\alpha _{3}=0$ we have the solution $\alpha _{1}^i-\kappa _{2}^{i}=\pi $ and
for $K_{1}^i=+1$ we have $\alpha _{1}^i-\kappa _{2}^{i}=0$.

So in summary, for both cases we can deduce that the two measurement directions are 
$\pi $ out of phase with each other, and for the first case ($\alpha
_{3}^i \neq 0$) we can freely vary $\alpha _{2}^i $ and $\alpha _{3}^i $, and for the
second case ($\alpha _{3}^i=0$), we can freely vary $\alpha_{1}^i $ and $\alpha
_{2}^i $ to change the initial quantum quantum state without affecting the game
Nash equilibrium~(NE) or payoffs \citep{Rasmusen,Binmore}. These results can be shown to imply in both cases that  $ \Omega = 0 $.

The associated payoff for Alice therefore becomes 
\begin{eqnarray}
\Pi _{A}(x^1,x^2,\dots x^N) & = & a_{00 \dots 0}  -\cos \gamma  \sum_{r=1}^{\left\lfloor (N+1)/2 \right \rfloor} C_{2r-1}^N[ a^{i0} (1-2 x^i ) +  a^{0i} ( 1 -2 x^i) ] 
\\ \nonumber
&&+  \sum_{r=1}^{\left\lfloor N/2 \right \rfloor}  C_{2r}^N [a^{1i} (1-2x^1)(1- 2 x^i)+ a^{0ij}(1-2 x^i)(1-2x^j)] .
\label{eq:AlicePayoffQuantumGameClassical}
\end{eqnarray}%

For example, for three players this will reduce to
\begin{eqnarray} \label{eq:AlicePayoffQuantumGameClassicalThreePlayer}
& & \Pi _{A}(x^1,x^2,x^3) \\ \nonumber
&=& a_{000} + a_{011} (1-2 x^2)(1-2 x^3)+ a_{110}(1-2x^1)((1- 2 x^2)+ (1-2x^3)) \notag \\
&-&\cos \gamma \left ( a_{111}(1-2x^1)(1-2x^2)(1-2x^3) + a_{100} (1 - 2 x^1 ) + a_{001}(2 - 2 x^2 -2 x^3)  \right ) , \nonumber
\end{eqnarray}%
in agreement with previous results for three-player games \citep{Chappell3Player}.
Now, we can write the equations governing the NE for the first player as 
\begin{eqnarray} \label{eq:NEEPREmbeddedStagHuntThree}
&&\Pi _{A}(x^{i \ast },x^{2 \ast }, \dots x^{N \ast })-\Pi _{A}(x^i,x^{2 \ast },\dots ,x^{N \ast }) 
\notag \\
&=&  ( x^{1*} - x^1 ) \left ( - \sum_{r=1}^{\left\lfloor N/2 \right \rfloor}  C_{2r}^N ( a^{1i} I^1 (1- 2 x^{i*})) +\cos \gamma  \sum_{r=1}^{\left\lfloor (N+1)/2 \right \rfloor} C_{2r-1}^N ( a^{i0} I^1 (1- 2 x^{i*}) ) \right ) \geq 0 . \notag 
\end{eqnarray}%
We are using $ I^1 $ as a placeholder, which has a value one, but ensures that the correct number of terms are formed from $ C_r^N () $.
For example, for three players we find the NE governed by 
\begin{eqnarray} \label{eq:NEEPREmbeddedStagHuntThreePlayer}
&&\Pi _{A}(x^{1\ast },x^{2\ast },x^{3\ast })-\Pi _{A}(x^1,x^{2\ast },x^{3\ast }) 
 \\
&=&(x^{1\ast }-x^1)[a_{110}(2x^{2 \ast }-1)+a_{101}(2x^{3 \ast }-1)+\cos \gamma
\{a_{100}+a_{111}(2x^{2\ast }-1)(2x^{3 \ast }-1)\}]\geq 0 , \notag
\end{eqnarray}%
in agreement with previous results \citep{Chappell3Player}.

\subsubsection{Symmetric game}

For a symmetric game we have $ a^{1 \dots 1} = b^{1 \dots 1} = \rm{etc} $, $ a^{0 \dots 0} =b^{0 \dots 0} = \rm{etc} $ and $ a_{11000 \dots 0} = a_{10100 \dots 0} = a_{10010 \dots 0} = \dots $, and similarly for other symmetries, and using these conditions for a symmetric game, we can find the NE for other players, such as Bob, from the constraint
\bea \label{eq:NEEPREmbeddedThreeReduced}
&&\Pi _{A}(x^{i \ast },x^{2 \ast }, \dots x^{N \ast })-\Pi _{A}(x^{i*},x^{2 },\dots ,x^{N \ast }) 
 \\
&=&  ( x^{2*} - x^2 ) \left (-\sum_{r=1}^{\left\lfloor N/2 \right \rfloor}  C_{2r}^N ( a^{1i} I^2 (1- 2 x^{i*})) +\cos \gamma  \sum_{r=1}^{\left\lfloor (N+1)/2 \right \rfloor} C_{2r-1}^N ( a^{i0} I^2 (1- 2 x^{i*}) ) \right ) \geq 0 .  \notag
\eea
We can see that the new quantum behavior is governed solely by the payoff
matrix and by the entanglement
angle $\gamma $, and not by other properties of the quantum state.

\subsubsection{Linear payoff relations}

We can see that as $ N \rightarrow \infty $, that we need to define an infinite number of components of the payoff matrix as shown by Eq.~(\ref{aDefn}).  Hence in order to proceed to solve specific games for large $ N $, we need to write the payoff matrix as some functional form of the measurement outcomes, as shown in Eq.~(\ref{eq:AlicePayoff}).
The simplest approach is to define linear functions over the set of player choices, as developed in \cite{FlitneyHollenberg2006}, defining the following general payoff function
\bea \label{eq:nonFlippingPlayerEqn}
\$_0 & = &  a n + b , \,\,\,\, \$_1  =  c n + d , 
\eea
where $ \$_0 $ is the payoff for players which choose their first measurement direction and $ \$_1 $ is the payoff for the players which choose their second measurement direction, and  where $ n $ is the number of players choosing their first direction and $ a,b,c,d \in \Re $.  

This approach enables us to simply define various common games.  For example the prisoner dilemma (PD), which has the essential feature that a defecting player achieves a higher payoff, is represented if we have $ c \ge a $, $ d > a + b $ and $ a>0 $.  These conditions ensure that if a cooperating player decides to defect, then his payoff rises as determined by Eq.~(\ref{eq:nonFlippingPlayerEqn}).
For example for $ a =3, b=-3,c=4,d=1 $ we have defined an $ N $ player PD, and for $ N =2 $ we find
\be \label{StdPDPayoffMatrix}
G_{ij}^A =\begin{bmatrix} 3 & 0 \\
5 & 1 \end{bmatrix} ,
\ee
which gives us the typical payoff matrix for two-player PD game.  In the EPR setting for the quantum game, a cooperating player is defined as the player who chooses their first measurement direction and a defecting player as one who chooses their second measurement direction.

For the Chicken game (also called the hawk-dove game) \cite{Rasmusen}, which involves the situation where the player that does not yield to the other is rewarded, but if neither player yields then they are both severely penalized, in this case we require $ c \ge a $ , $ d < a + b $ and $ a > 0 $ and for the minority game, an implementation would be $ c = -a $, $ a < 0 $ and $ d = b + a N $ which rewards a minority choice and punishes a majority one.  
Hence we are led to define 
\be \label{defnp1p2}
 p_1 = d-(a+b)  , \,\,\,\ p_2 = c - a ,
\ee
as two key determinants of quantum games, and we will find that the NE is indeed a function of $p_1 $ and $ p_2 $ alone, see Eq.~(\ref{eq:NEConditionsGHZ}).  With this definition the PD game is selected if $p_1>0$ and $ p_2 \ge 0$ and the minority game with $ p_1<0 $ and $ p_2 > 0$ for example.

It should be noted that while the definition in Eq.~(\ref{eq:nonFlippingPlayerEqn}) can generally define an infinite set of PD games through simply putting conditions on $ p_1 $ and $ p_2 $, it is still only a subset of the space of all possible PD games defined over $ N \times N $ payoff matrices. 

Using the linear functions defined in Eq.~(\ref{eq:nonFlippingPlayerEqn}) we find
\bea \label{aseriesCalc}
a^{0 \dots 0} & = & \frac{1}{4} ( N(c+a)-p_2 + 2(b+d)) \\ \nonumber
a^{10 \dots 0} & = & -\frac{1}{4} ( (N-1)(c-a) + 2(d-(a+b))) = -\frac{1}{4} \left ( (N-1)p_2 + 2 p_1 \right )\\ \nonumber
a^{110 \dots 0} & = & -\frac{c-a}{4} = -\frac{p_2}{4} \\ \nonumber
a^{1110 \dots 0} , a^{11110 \dots 0} \dots & = & 0  \nonumber
\eea
and
\bea \label{aseriesCalc0Type}
a^{010 \dots 0} & = & \frac{c+a}{4} \\ \nonumber
a^{011 \dots 0} , a^{0111 \dots 0} , \dots & = & 0.  \nonumber
\eea
If required, Eq.~(\ref{eq:nonFlippingPlayerEqn}) can be extended with quadratic terms in $ n $ to allow a greater variety of PD games to be defined, and we find that if this is done that one extra term is added to the series in Eq.~(\ref{aseriesCalc}) and Eq.~(\ref{aseriesCalc0Type}). 

\subsubsection{NE and payoff for linear payoff relations}

We can see that the series in Eq.~(\ref{aseriesCalc}) and Eq.~(\ref{aseriesCalc0Type}) terminates, which thus allows us to simplify the NE conditions, for the first player to
\be \label{eq:NEConditionsGHZ}
  ( x^{1*} - x^1 ) \left (p_2 \sum_{i=2}^N (1- 2 x^{i*}) -\cos \gamma  \left ( (N-1) p_2+ 2 p_1 \right )  \right ) \geq 0  
\ee
and similarly for the other $ N-1$ players, which thus determines the available NE for all games, defined as linear functions, in terms of the two parameters $ p_1 $ and $ p_2 $.

The payoff can then also be simplified for the first player to
\bea \label{payoffFirstPlayer}
\Pi_A  & = & \frac{1}{4} \Big ( 2(b+d) -p_2 + (c+a) \big (N - \cos \gamma \sum_{i=2}^N (1- 2 x^{i}) \big )  \\ \nonumber
&+ &(1-2 x^1) \big (\cos \gamma  ((N-1)p_2 +2 p_1) - p_2 \sum_{i=2}^N (1- 2 x^{i}) \big ) \Big ) . \nonumber
\eea
For the minority game defined previously, we find $ (N-1) p_2+ 2 p_1 = 0 $, which gives an interesting result for this game that both the NE and the payoff are unaffected by the entanglement of the state.

\subsubsection{Prisoner dilemma (PD)}

For the PD, having $p_2 \ge 0$ and $p_1>0$, and we find from the equation for Nash equilibrium in Eq.~(\ref{eq:NEConditionsGHZ}) that in order to produce the classical outcome we require $ \sum_{i=2}^N (1- 2 x^{i*}) < \cos \gamma ( N - 1 + 2 p_1/p_2 ) $ which thus requires $  \cos \gamma > \frac{N-1}{N-1+2 p_1/p_2} $  and hence the phase transitions, in terms of $ \cos \gamma $, are given by
\be  \label{eq:PDTransitions}
 \frac{N-1-2n}{N-1+\delta} < \cos \gamma < \frac{N+1-2n}{N-1+\delta} = \lambda_n ,
\ee
where $ \delta  = \frac{2 p_1 }{p_2}  $, and with the PD $ \delta \in (0,\infty) $, and hence the above inequality will hold for $ N \ge 2 $.  
So in summary, at the classical limit we have all players defecting, and then we have the transition to the non-classical region at $ \lambda_1 $ and we then have equally spaced transitions as entanglement increases down to maximum entanglement where we have the number of players cooperating $ n = \lfloor N/2 \rfloor $.  That is, we always have the same number of transitions for a given number of players, but they concertina closer together as the first transition $ \lambda_1 $, moves towards zero, through changing the game parameters, $ p_1 $ and $ p_2 $.

\setlength{\unitlength}{1mm}
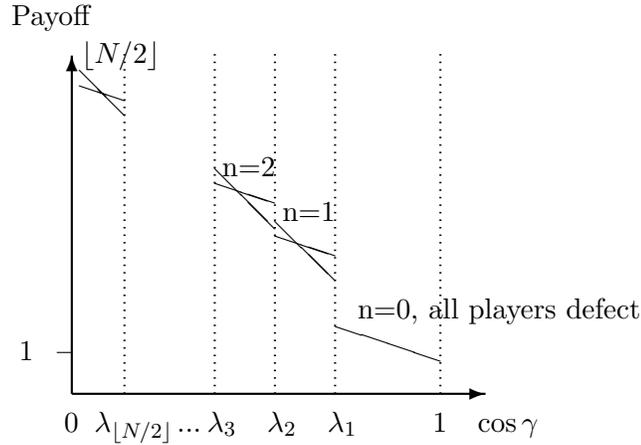
\begin{figure}[htb]
\begin{picture}(75,80)
\put(11,48){\line(3,-1){6}}
\put(11,50){\line(1,-1){6}}
\put(29,35){\line(3,-1){8}}
\put(29,37){\line(1,-1){8}}
\put(37,28){\line(3,-1){8}}
\put(37,30){\line(1,-1){8}}
\put(45,16){\line(3,-1){14}}
\thinlines
\put(8,12.5){\line(1,0){2}}

\thicklines 
\put(10,7){\vector(1,0){55}} \put(10,7){\vector(0,1){45}}

\put(2,56){Payoff} \put(64,2){$ \cos \gamma $} 
\put(9,2){$ 0 $} \put(58,2){$ 1 $}
\put(3,11.5){$ 1 $} \put(44,2){$ \lambda_1 $} \put(36,2){$ \lambda_2 $}  \put(28,2){$ \lambda_3 $} 
\put(13,2){$ \lambda_{\lfloor N/2 \rfloor} $} \put(24,2){$ ... $} 
\put(10,51){ $ \lfloor N/2 \rfloor$} \put(30,36){n=2}\put(38,30){n=1}\put(48,17){n=0, all players defect}
\multiput(17,7)(0,1){46} {\line(0,1){0.2}} 
\multiput(29,7)(0,1){46} {\line(0,1){0.2}} 
\multiput(37,7)(0,1){46} {\line(0,1){0.2}} 
\multiput(45,7)(0,1){46} {\line(0,1){0.2}} 
\multiput(59,7)(0,1){46} {\line(0,1){0.2}} 
\end{picture}
\caption{Phase structure for $N $-player Prisoner dilemma.  For $ \cos \gamma > \lambda_1 $ we identify the classical regime, where all players defect, and as entanglement increases we find an increasing number of players cooperating, up to $ \left \lfloor N/2 \right \rfloor $ near maximum entanglement.  The left and right hand edges of the boundaries each form an inverted parabola in $ \cos \gamma $ given by Eq.~(\ref{PDExampleParabolic}). \label{GA3Space}}
\end{figure}

The maximum payoff, close to maximum entanglement, can be found from Eq.~(\ref{payoffFirstPlayer}) as 
\bea
\Pi_A^c & = & \frac{1}{4}(2(b+d)+(c+a)N+(c-a)_{N \in {\rm{Odd}}}) \\ \nonumber
\Pi_A^d & = & \frac{1}{4}(2(b+d)+(c+a)N-(c-a)_{N \in {\rm{Odd}}}) , \nonumber
\eea
where the final $ (c-a) $ term only occurs for odd $ N $.  So for $ N $ even the payoffs are equal, but for odd $ N $, the cooperating player receives a higher or equal payoff to the defecting player.
The payoff rises linearly with $ N $, whereas without entanglement, we have the payoff fixed at $d$ units from Eq.~(\ref{eq:nonFlippingPlayerEqn}).

\subsubsection{The conventional prisoner dilemma (PD) game for all $ N $}

For the special case with the PD settings shown in Eq.~(\ref{StdPDPayoffMatrix}), which gives the conventional PD game for two players, we find from Eq.~(\ref{defnp1p2}), $ p_1 = 1 $ and $ p_2 = 1 $, and so we can then simplify the general NE conditions in Eq.~(\ref{eq:NEConditionsGHZ}), for the first player to
\be \label{eq:NEConditionsGHZPDExample}
  ( x^{1*} - x^1 ) \left (\sum_{i=2}^N (1- 2 x^{i*}) -(N+1) \cos \gamma \right ) \geq 0  
\ee
and similarly for the other $ N-1$ players.
The left and right edges of each NE zone, shown in Fig.~\ref{GA3Space}, can now be written from Eq.~(\ref{eq:PDTransitions}) as
\be  \label{eq:PDTransitionsExample}
 \frac{N-1-2n}{N+1} < \cos \gamma < \frac{N+1-2n}{N+1} .
\ee
In each zone we find the payoff for cooperation and defection, from Eq.~(\ref{payoffFirstPlayer}), now given by
\bea \label{payoffPlayerExample}
\Pi^c & = & \frac{1}{2} \left ( 4 N -2 - n - ( 4 + 4 N -7 n) \cos \gamma  \right ) \\  \nonumber
\Pi^d & = & \frac{1}{2} \left ( 3 N -2 + n + ( 4 - 3 N+ 7 n) \cos \gamma  \right ) , \nonumber
\eea
which defines the payoff diagram for an $ N $ player PD, and which produces the classical PD at $ N =2 $ at zero entanglement.

At each left hand boundary, for the defecting player, we have from Eq.~(\ref{eq:PDTransitionsExample}), $ \frac{N-1-2 n}{N+1} = \cos \gamma $ or $ n = \frac{1}{2} \left ( N -1 - (N+1) \cos \gamma \right ) $.  Substituting this into the defecting player payoff in Eq.~(\ref{payoffPlayerExample}), we find
\be \label{PDExampleParabolic}
\Pi^d = -3 + \frac{7}{4} (N+1) (1 - \cos^2 \gamma) = -3 + \frac{7}{4} (N+1) \sin^2 \gamma ,
\ee
for the defecting players' payoff.
We thus see that the payoff at each boundary follows a downwards parabolic curve in $ \cos \gamma $, if drawn on Fig.~\ref{GA3Space}.  If we allow $ N $ to increase without limit, then the boundaries would concertina infinitesimally close together, and in the limit as $ N \rightarrow \infty $, the payoff's would form a continuous downward parabolic curve in $ \cos \gamma $ given by Eq.~(\ref{PDExampleParabolic}).  The special case of the PD selected here with $ p_1 = 1 $ and $ p_2 = 1 $ forms a parabola, whereas for the general case of a PD game with $p_2 \ge 0$ and $p_1>0$ from Eq.~(\ref{defnp1p2}), we will produce a quadratic curve in $ \cos \gamma $ for the payoff.  We can also see that this will be a general feature for all games defined using linear functions as both the NE in Eq.~(\ref{eq:NEConditionsGHZ}) and the payoffs in Eq.~(\ref{payoffFirstPlayer}) are linear in $ \cos \gamma $, therefore typically producing a payoff diagram quadratic in $ \cos \gamma $.

We can also note that Eq.~(\ref{payoffPlayerExample}) indicates a different payoff for the defecting and cooperating player at the NE.  If a player decides to try to change their choice in order to improve their payoff, often a lower payoff will be the outcome, because overall the player's choices have now moved away from the NE.  This then illustrates the value of coalitions and in aligning one's choices with the coalition with the higher payoff \citep{IqbalToor3,FlitneyGreentree}.

\subsection{W entangled state}

Following the same procedure as used for the GHZ-type state, we find the probability distribution for the W-type state
\be \label{eq:finalDensityW}
P_{k^1 \dots k^N} = \frac{1}{N 2^{N}}  ( N  + \sum_{r=1}^{N} (N-2r) C_{r}^N (\epsilon^{i} K^i) +2 \sum_{r=2}^{N} C_{r}^N (\epsilon^{i} \epsilon^{j} \epsilon^{k} ( X_2^i X_2^j + X_1^i X_1^j) K^k ) ). 
\ee
We can then find the payoff function for the first player, Alice
\be \label{WtypePrisonerDilemma}
\Pi_A (\kappa^1,\dots ,\kappa^N) =  N a^{0 \dots 0} + \sum_{r=1}^{N} (N-2r) C_r^N (a^{i} K^i) +2 \sum_{r=2}^{N}  C_{r}^N  (a^{i j k} ( X_2^i X_2^j + X_1^i X_1^j) K^k )   
\ee
and similarly for other players.
However with the W-type state it is impossible to turn off the entanglement, and so it will not be possible to embed the classical game, as we have done with the GHZ-type state.  Hence we will not proceed any further except to show the result of maximizing the payoff function in Eq.~(\ref{WtypePrisonerDilemma}) for the PD.

\subsubsection{Prisoner Dilemma (PD)}

For the PD we can maximize the payoff function, and we find that we require all players to defect, for all $N$ and the resultant payoff for the first player Alice and hence all players is 
\be
\Pi_A = c+d - \frac{c+d-(a+b)}{N}.
\ee
So as $ N \rightarrow \infty $, then the payoff approaches $ c + d $ from below.

\section{Conclusion}

Using Clifford's geometric algebra, the probability distribution is found for general measurement directions on a general $ N $ qubit entangled state, for the GHZ-type state shown in Eq.~(\ref{eq:finalDensityGeneral}) and for the W-type state shown in Eq.~(\ref{eq:finalDensityW}).

Linear functions parameterized by the number of players selecting their first measurement direction for an $ N $ player game are then defined as shown in Eq.~(\ref{eq:nonFlippingPlayerEqn}), from which games can then be easily defined for general $ N $. Using these linear functions, the Nash equilibrium and payoff relations are then determined for general $ N $ as shown in Eq.~(\ref{eq:NEConditionsGHZ}) and Eq.~(\ref{payoffFirstPlayer}) respectively.  We also find a general feature for these games of producing a payoff diagram with phase transition boundaries quadratic in $ \cos \gamma $, as shown in Fig.~\ref{GA3Space}.  If the linear functions are increased in order, then we would expect the payoff diagram to become a higher order polynomial in $ \cos \gamma $.

As a specific example the PD is solved for a general $ N $ and we find an interesting feature, that the payoffs at the Nash equilibrium are equal for the defecting and cooperating player only for even $N$ and also in the limit of large $ N $ the payoff rises linearly with $ N $ given by $ (c+a) N/4 $ for the GHZ-type state.

At maximum entanglement the payoff for the GHZ-type and W-type states for the PD become equal at $ N = 2 $, producing the formula from the parameters of the linear functions as
\be
\Pi_{\rm{GHZ}} =  \Pi_{\rm{W}} = \frac{a+b+c+d}{2} .
\ee
This equality is to be expected at $ N = 2 $, because these two states are equivalent under local operations.

In summary, we have produced a general quantum game environment, with the number of players $ N \ge 2 $, which will embed the classical game at zero entanglement, and using linear functions we determine the NE and player payoffs for general $ N $. These general results thus subsume previous analyses for two-player and three-player games in an EPR setting \citep{ChappellB,Chappell3Player}.

\newpage

\appendix

\section{Calculating the observables}

The following three results are useful when calculating
the observables, for example as in Eq.~(\ref{eq:Method1MeasureE}), for a given measurement direction $\kappa $. If we have a
rotor defined as 
\begin{equation}
R=\mathrm{e}^{-\alpha _{3}{\iota }\sig _{3}/2}\mathrm{e}^{-\alpha _{1}{%
\iota }\sig _{2}/2}\mathrm{e}^{-\alpha _{2}{\iota }\sig _{3}/2},
\end{equation}%
then we find the following results 
\begin{subequations}
\begin{align}
\langle {\iota }R\sig _{3}R^{\dagger }{\iota }\sig _{3}\mathrm{e}%
^{\kappa {\iota }\sig _{2}}\rangle _{0}& =-\cos \alpha _{1}\cos \kappa
-\cos \alpha _{3}\sin \alpha _{1}\sin \kappa ,  \notag \\
\langle {\iota }R\sig _{2}R^{\dagger }{\iota }\sig _{3}\mathrm{e}%
^{\kappa {\iota }\sig _{2}}\rangle _{0}& =\sin \kappa (\cos \alpha
_{2}\sin \alpha _{3}+\sin \alpha _{2}\cos \alpha _{3}\cos \alpha _{1})-\sin
\alpha _{1}\sin \alpha _{2}\cos \kappa ,  \notag \\
\langle {\iota }R\sig _{1}R^{\dagger }{\iota }\sig _{3}\mathrm{e}%
^{\kappa {\iota }\sig _{2}}\rangle _{0}& =-\sin \kappa (\cos \alpha
_{1}\cos \alpha _{2}\cos \alpha _{3}-\sin \alpha _{2}\sin \alpha _{3})+\sin
\alpha _{1}\cos \alpha _{2}\cos \kappa .  \notag 
&
\end{align}
\end{subequations}

\section{Prisoner Dilemma, modified linear functions}

Flitney and Hollenberg \citep{FlitneyHollenberg2006}, define linear functions slightly differently from Eq.~(\ref{eq:nonFlippingPlayerEqn}), with a special case at $ m = 1$.  The advantage of their definition is that the phase diagram has entanglement transitions that are independent of $ N $, but with the disadvantage that we need to administer this special case in the calculations.

They define the following general payoff function, for the cooperating player
\bea
\$_C & = &  0 \,\,\,\,\,\,\,\,\,\,\,\,\, \,\,\,\,\,\,\,\,\,\,\,\,\,\,\,\,\,\,\,\,\,\, {\rm{if}} \,\, {\rm{m}} =1 \\ \nonumber
     & = &  3 + 4 (m-2) \,\,\,\,\, {\rm{if}} \,\, {\rm{m}}>1  \nonumber
\eea
and for the defecting player
\be
\$_D =  5 + 4 (m-1) ,
\ee
where $ m $ is the number of players cooperating.
Using the definitions above we find
\bea
a^{0 \dots 0} & = & 2(N-1) + 1/2^N \\ \nonumber
a^{10 \dots 0} & = & -1 + 1/2^N \\ \nonumber
a^{11 \dots 0} & = & -1/2^N \\ \nonumber
a^{111 \dots 0} & = & 1/2^N \\ \nonumber
a^{1111 \dots 0} & = & -1/2^N \\ \nonumber
a^{11111 \dots 0} & = & 1/2^N  \nonumber
\eea
and
\bea
a^{010 \dots 0} & = & 2 - 1/2^N \\ \nonumber
a^{011 \dots 0} & = & 1/2^N \\ \nonumber
a^{0111 \dots 0} & = & -1/2^N \\ \nonumber
a^{01111 \dots 0} & = & 1/2^N \\ \nonumber
a^{011111 \dots 0} & = & -1/2^N,  \nonumber
\eea
which are quite different from those obtained from the GHZ-type state in Eq.~(\ref{aseriesCalc}) as these do not terminate, and hence we will indeed find a different phase diagram.  Our approach is preferred because the series terminates, allowing the general case to be more easily handled.

\bibliographystyle{plain}
\bibliography{quantum}

\end{document}